\documentclass[sigconf]{acmart}

\usepackage{booktabs} 
\usepackage{subcaption}

\copyrightyear{2018} 
\acmYear{2018} 
\setcopyright{acmcopyright}
\acmConference[RecSys '18]{Twelfth ACM Conference on Recommender Systems}{October 2--7, 2018}{Vancouver, BC, Canada}
\acmBooktitle{Twelfth ACM Conference on Recommender Systems (RecSys '18), October 2--7, 2018, Vancouver, BC, Canada}
\acmPrice{15.00}
\acmDOI{10.1145/3240323.3240377}
\acmISBN{978-1-4503-5901-6/18/10}

\begin{document}
\title{Word2vec applied to Recommendation: Hyperparameters Matter}

\author{Hugo Caselles-Dupr{\'e}\textsuperscript{1}\textsuperscript{2}}
\authornote{Work done while interning at Deezer}
\affiliation{%
  \institution{\textsuperscript{1} Flowers Laboratory (ENSTA ParisTech \& INRIA)}
  \institution{\textsuperscript{2} Softbank Robotics Europe}
  \city{Paris} 
  \country{France} 
  \postcode{75009}
}
\email{caselles@ensta.fr}

\author{Florian Lesaint}
\affiliation{%
  \institution{Deezer SA}
  \streetaddress{12 rue d'Ath\e nes}
  \city{Paris} 
  \country{France} 
  \postcode{75009}
}
\email{flesaint@deezer.com}

\author{Jimena Royo-Letelier}
\affiliation{%
  \institution{Deezer SA}
  \streetaddress{12 rue d'Ath\e nes}
  \city{Paris} 
  \country{France}}
\email{jroyo@deezer.com}

\renewcommand{\shortauthors}{Caselles-Dupr\'e et al.}

\begin{abstract}

Skip-gram with negative sampling, a popular variant of Word2vec originally designed and tuned to create word embeddings for Natural Language Processing, has been used to create item embeddings with successful applications in recommendation. While these fields do not share the same type of data, neither evaluate on the same tasks, recommendation applications tend to use the same already tuned hyperparameters values, even if optimal hyperparameters values are often known to be data and task dependent. We thus investigate the marginal importance of each hyperparameter in a recommendation setting through large hyperparameter grid searches on various datasets. Results reveal that optimizing neglected hyperparameters, namely negative sampling distribution, number of epochs, subsampling parameter and window-size, significantly improves performance on a recommendation task, and can increase it by an order of magnitude. Importantly, we find that optimal hyperparameters configurations for Natural Language Processing tasks and Recommendation tasks are noticeably different.

\end{abstract}

%
%


\maketitle

\section{Introduction}

Word2vec (W2V) methods \cite{word2vec1,word2vec2} come from the Natural Language Processing (NLP) community. They were designed to produce low-dimensional distributional word representations. They were successfully applied to recommendation \cite{prod2vec} to generate user and product embeddings as they can scale to millions of items.

Word corpora and sequences of items are two radically different type of data. Text data has a particular linguistic structure \cite{bybee2001frequency}, constrained by grammatical and conjugating rules. Sequences of items, such as listening sessions or e-commerce purchase histories, have a different structure induced by the user's behaviour and the items' nature \cite{greer1973adult}. Moreover, linguistic and recommendation tasks are different. Intuitively, having accurate embeddings for popular items is crucial to perform on recommendation related tasks \cite{steck2011item}, where top items often represent most of the content users interacted with. On the contrary, for linguistic tasks, frequent words, mostly linking words, are not relevant \cite{word2vec2} and so are their embeddings. Since hyperparameters choices are generally known to be data and task dependent \cite{hutter2014efficient}, we expect optimal hyperparameter configuration to be different for NLP and recommendation. 

W2V methods depend on several hyperparameters, some of which are already tuned to some extent by the algorithms' designers in order to perform well for NLP tasks such as word similarity and analogy detection \cite{levy2015improving}, such that most renowned implementations \cite{gensim, meng2016mllib} set these values as default. In previous work that used W2V for recommendation \cite{prod2vec,barkan2016item2vec,musto2015word,ozsoy2016word,metaprod2vec,nedelec2016content2vec}, the values of these hyperparameters are rarely discussed.

Thus, we study the marginal importance of each hyperparameter of Skip-gram with negative sampling (SGNS) in a recommendation setting, using Next Event Prediction (NEP) as an offline proxy for a recommendation task. We perform large hyperparameter grid searches on four different types of recommendation datasets (two of music, one of e-commerce and one of click-stream). This allows us to identify four hyperparameters, namely negative sampling distribution, number of epochs, subsampling parameter and window-size, which can significantly improve performance on the NEP task. This confirms that optimal values for these hyperparameters are data and task dependent, and that best configurations for recommendation tasks are radically different than those for NLP tasks, especially regarding the negative sampling distribution. 

We first describe W2V methods and associated hyperparameters in Section~2. Then, we present the experiments in Section~3 and results in Section~4 before concluding in Section~5.

\section{Word2vec}
\subsection{Methods}
W2V \cite{word2vec1,word2vec2} is a group of word embedding algorithms that provides state-of-the-art results on various linguistic tasks \cite{levy2015improving}. They are based on the Distributional Hypothesis \cite{sahlgren2008distributional}, which states that words that appear in the same contexts tend to purport similar meanings. The most common method, SGNS, used in the remaining of the paper, seeks to represent each word $w$ as d-dimensional vector $\vec{w}$, such that words that are similar to each other have similar vector representations. It does so by maximizing a function of products $\vec{w} \cdot \vec{c}$ where $c$ appears in the context of $w$ (a window around $w$ of maximum size $L$), and minimizing the same function for negative examples $(w, c_N)$ where $c_N$ does not necessarily appear in the  context of $w$. The loss function is 

\begin{equation} 
\label{eq3}
\ell_{\text{sgns}} = - \log(\sigma(\vec{w}  \cdot  \vec{c})) - \sum_{i=1}^{k} \log(\sigma(-\vec{w}  \cdot  \vec{c}_{N,i})) \,, \end{equation}

\noindent where $\sigma$ is the sigmoid function. For each observation of $(w, c)$, SGNS forms $k$ negative examples $(w, c_{N,i})_{i \in \{1,..,k\}}$ by sampling (hence the term "negative sampling") $k$ words in the corpus from a $\alpha$-smoothed unigram distribution:
\begin{equation} 
\label{eq1}
  P(c)=\frac{f(c)^{\alpha}}{\sum_{c'} f(c')^{\alpha}} \,.
\end{equation}
Here, $f(c)$ represents the frequency of the word $c$ and the parameter $\alpha \in \mathbb{R}$ smoothes the distribution. An $\alpha$ equal to 1 leads to a sampling based on the frequency distribution, an $\alpha$ equal to 0 makes items being sampled equally, while a negative $\alpha$ makes unpopular items being sampled more often than popular onces. The parameter $\alpha$ is empirically set to $0.75$ following \cite{word2vec2}.
Better performance and faster training can be obtained by using a dynamic window-size (i.e.: randomly sampling the window size between 1 and $L$) or by randomly removing words  (sub-sampling) that are more frequent than some threshold $t$ with a probability 
\begin{equation} 
\label{eq2}
  p(c)=\frac{f(c)-t}{f(c)} - \sqrt{\frac{t}{f(c)}}.
\end{equation}

In recommendation settings, such as music consumption or online shopping, a revised version of the Distributional Hypothesis is adopted to justify the use of SGNS, stating that items that appear in the same contexts share similarities. \citeauthor{prod2vec} \cite{prod2vec} proposed the use of SGNS on sequences of items to form item embeddings employed in recommendation applications. W2V-based item embeddings have since been successfully used in numerous recommendation scenarios \cite{barkan2016item2vec,metaprod2vec,ozsoy2016word,musto2015word}. Since then, this method has been derived to handle problems specific to recommendation. For example, Meta-Prod2vec \cite{metaprod2vec}, improves upon Prod2vec by using the item meta-data side information to regularize the final item embedding, and authors show that they outperforms Prod2vec on NEP for music, globally and especially in a cold-start regime.

\subsection{Hyperparameters}

In the following, we describe the role and classically used values of the hyperparameters in the investigated literature \cite{prod2vec,barkan2016item2vec,musto2015word,ozsoy2016word,metaprod2vec,nedelec2016content2vec}, for which simultaneous optimization significantly improved NEP performance.


Negative pairs of items $(w,c_N)$ are sampled from the negative sampling distribution which is parametrized by $\alpha$ in Equation \eqref{eq1}. The original smoothed unigram distribution, proposed in \cite{word2vec2}, samples items proportionally to their frequency raised to the power $\alpha=0.75$. This value was empirically chosen because it outperformed the uniform ($\alpha=1$) and unigram ($\alpha=0$) distributions on every linguistic task tested by the authors. This result was further confirmed in \cite{levy2015improving}, where the authors extensively studied the marginal effect of optimizing each hyperparameter of W2V. Consequently, widely used implementations of W2V (e.g. Gensim \cite{gensim}) use this value by default, and does not present it as tunable. We assume that works that do not discuss this parameter rely on its commonly accepted default value.

The number of epochs $n$ controls the total number of times SGNS goes over each item of the dataset, which has a direct impact on the duration and the quality of the training. Its default value is set to $5$ in Gensim \cite{gensim}. Some work did hyperparameter search on the number of epochs \cite{ozsoy2016word, metaprod2vec} on a range we extended in this work, or do not develop on the methods nor the final values used.

The window-size is sampled randomly between 1 and the maximum window-size $L$. It controls how wide the gap between two items in a sequence can be, such that they are still considered in the same context. The default value is set to $5$ in Gensim \cite{gensim}. Some authors claim that it is best to use a "infinite" window-size \cite{barkan2016item2vec}, meaning that the whole sessions is considered as one context, but most arbitrarily use a fixed value without further discussion.



Higher-frequency items are randomly sub-sampled, according to Equation \eqref{eq2}. The default value of the parameter $t$ is set to $10^{-3}$ in Gensim \cite{gensim}. This variable is hardly discussed in the investigated literature \cite{barkan2016item2vec,prod2vec}, and to our knowledge never optimized.

\section{Experiments}

We study the influence of 7 hyperparameters (including $n,L,t,\alpha$) on final performance by evaluating SGNS on a recommendation task based on items embeddings, with 4 recommendation datasets coming from diverse sources. Our code is available online. \footnote{Code and datasets for reproducing our results can be found at \url{https://github.com/deezer/w2v_reco_hyperparameters_matter}}

\begin{figure}[htb]
    \centering
    \subcaptionbox{30Music dataset}[.45\linewidth][c]{
    \includegraphics[width=.5\linewidth]{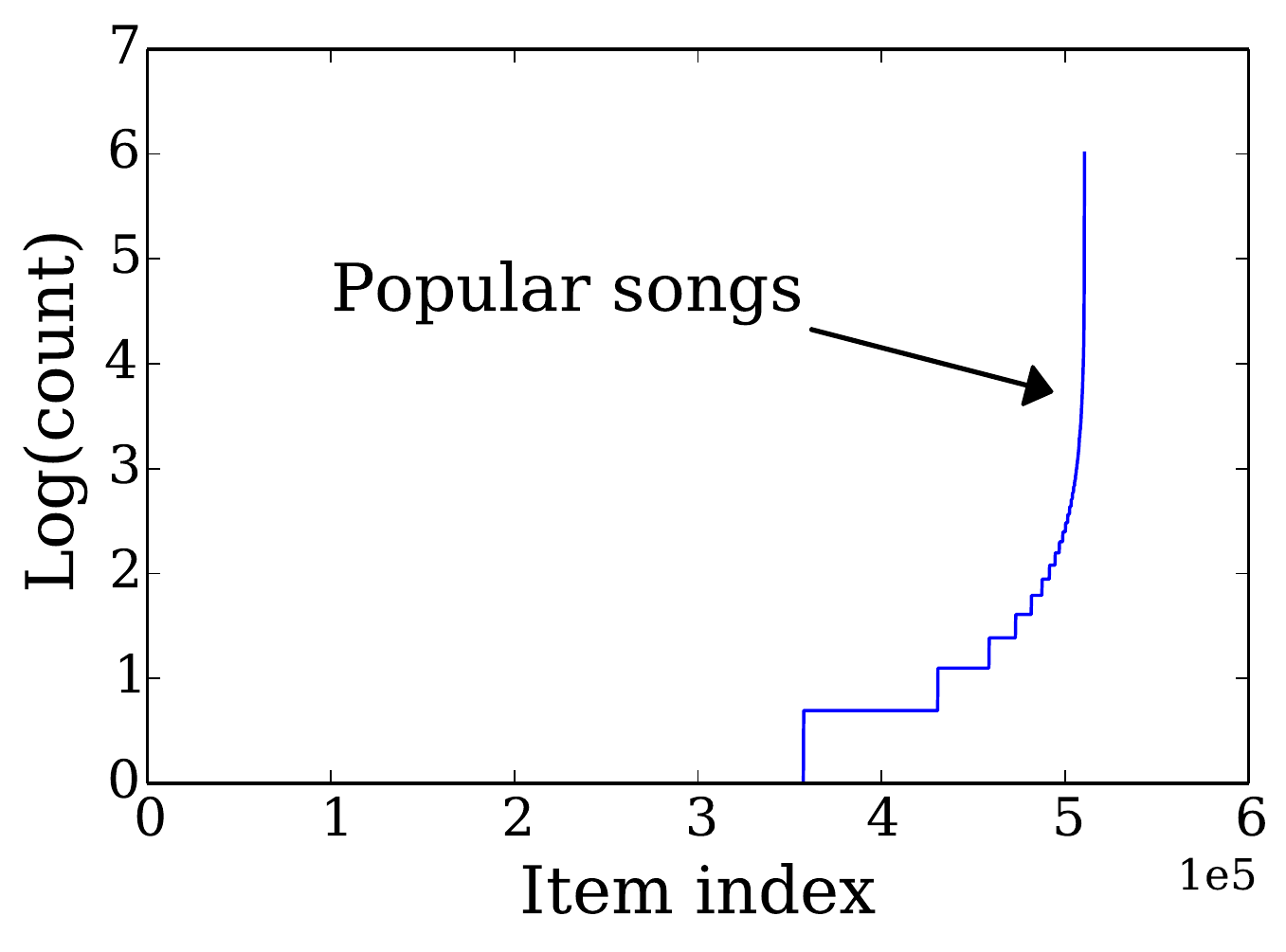}}\quad
    \subcaptionbox{Deezer dataset}[.45\linewidth][c]{
    \includegraphics[width=.5\linewidth]{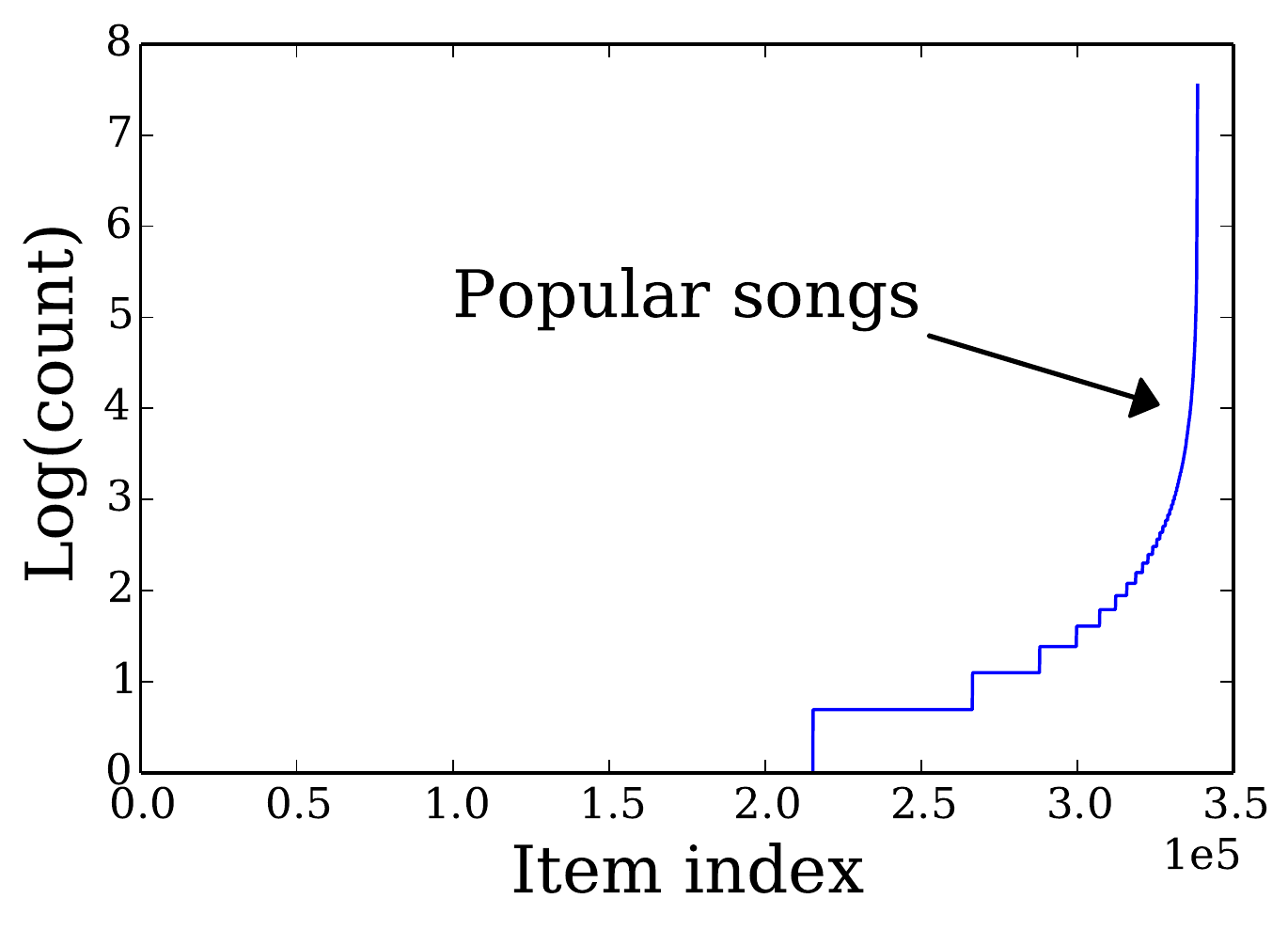}}\\
    \subcaptionbox{E-commerce dataset}[.45\linewidth][c]{
    \includegraphics[width=.5\linewidth]{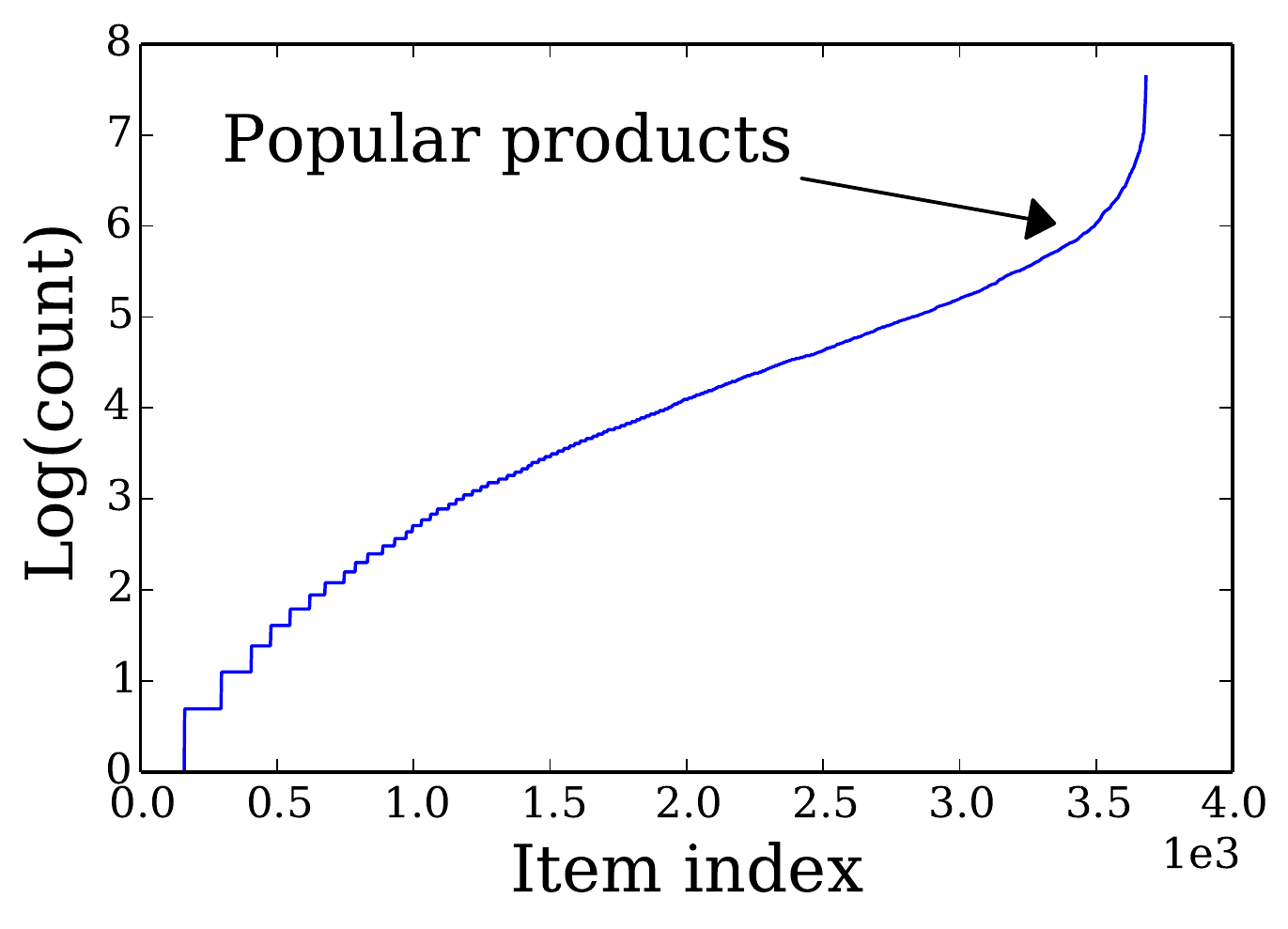}} \quad
    \subcaptionbox{Click-Stream dataset}[.45\linewidth][c]{
    \includegraphics[width=.5\linewidth]{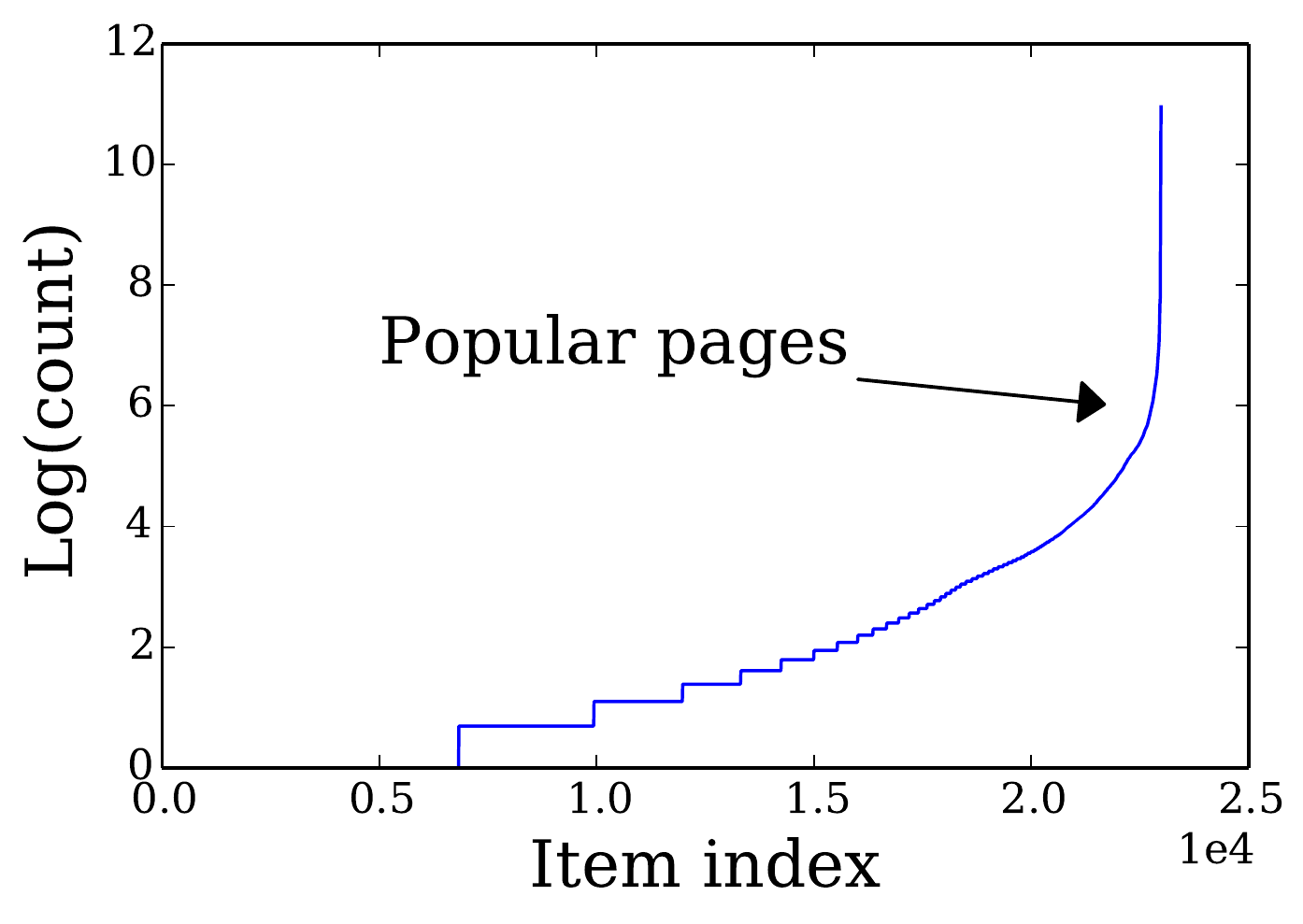}}
    \caption{Log count distributions of the considered datasets.}
\label{fig1}
\end{figure}

\subsection{Datasets} 
\label{subsec:Datasets}

\subsubsection{Music datasets}
We rely on 2 sets of listening sessions. The former, "30Music" \cite{turrin201530music}, composed of listening and playlists data retrieved from Internet radio stations, is open and commonly used for recommendation \cite{metaprod2vec,jannach2017leveraging,ben2017groove,brusamento2016explicit}. The latter is a dataset of listening sessions from Deezer, a French on-demand music streaming service. Both are composed of $100$k sessions sampled from the original datasets. We refer to these datasets as 30Music and Deezer datasets, respectively. Log count distributions for these datasets are shown in Figure \ref{fig1}. The log count distribution tail is sharp: there is an important discrepancy between popular and unpopular items. We notice a strong resemblance between the two distributions, which suggests similarities of music usage between users of the two platforms. 

\subsubsection{E-commerce dataset}
We use an open Online Retail dataset \cite{chen2012data} composed of transactions occurring between 01/12/2010 and 09/12/2011 for a UK-based and registered non-store online retail. It is composed of $4234$ user purchase histories. Compared to music data, the log count distribution tail is heavier: the discrepancy between popular and unpopular items is smaller. 

\subsubsection{Click-stream dataset}
We use the "kosarak" dataset \cite{bodon2003fast}, which contains anonymized click-stream data of a Hungarian on-line news portal. It is composed of $83625$ user click-stream histories. The log count distribution tail is comparable to the two music datasets.

\subsection{Task and metrics}

We evaluate the item embeddings on the NEP task, a common way to assess the quality of item embeddings \cite{metaprod2vec, letham2013sequential, rendle2010factorizing} for recommendation. We consider time ordered sequences of user interactions with the items. We split each sequence into training, validation and test sets. We first fit the SGNS model on the first $(n-1)$ elements of each user sequence; then, we use the performance on randomly sampled ($(n-1)$-th, $n$-th) pairs of items (validation set) to bench the hyperparameters, and finally, we report our final results by performing prediction on randomly sampled ($(n-1)$-th, $n$-th) pairs of items (test set, disjoint with validation set). For prediction, we use the last item in the training sequence as the query item and predict the $k$ closest items to the query item using a nearest-neighbor approach \cite{cover1967nearest}. We use $10$k test/validation pairs for 30Music, Deezer and Click-Stream datasets and $2$k for E-Commerce dataset. We evaluate with the following metrics:
\begin{itemize}
\item Hit ratio at K (HR@K). It is equal to $1$ if the test item appears in the list of $k$ predicted items and $0$ otherwise \cite{hr}.
\item Normalized Discounted Cumulative Gain (NDCG@K). It favors higher ranks in the ordered list of predicted items \cite{ndcg}. Since only one of the $k$ retrieved items can be relevant (i.e. equal to the $n$-th item in the sequence), the formula writes
\begin{equation}
\label{eq14}
NDGC@K = \left\{
    \begin{array}{ll}
        \frac{1}{log_2(j+1)} & \mbox{if $j^{th}$ predicted item is correct } \\
        0 & \mbox{otherwise}
    \end{array}
\right.
\end{equation}
\end{itemize}

\subsection{Experimental setup}

We use a modified implementation of Gensim \cite{gensim} for our experiments, such that parameter $\alpha$ (Eq. \eqref{eq3}) becomes tunable. We perform a hyperparameter search ($300$k models evaluated) on: the number of epochs $n$ ($10$ to $200$ with step of $+10$), the window-size $L$ ($3,7,12,15$), the sub-sampling parameter $t$ (Eq. \eqref{eq1}) ($10^{-5}$ to $10^{-1}$ with step of $\times 10$), the negative sampling distribution parameter $\alpha$ (Eq. \eqref{eq2}) ($-1.4$ to $1.4$ with step of $+0.2$), the embedding size ($50$ to $200$ with a step of $50$), the number of negative samples ($5$ to $20$ with a step of $5$) and the learning rate ($0.0025$ to $0.25$ with a step of $\times 10$). The marginal benefit of including the 3 latter variables to the optimization is not significant, with less than $2$\% in terms of performance. Thus, for readability, we only focus on the influence of the 4 first hyperparameters and keep the other fixed to default values (respectively $50, 5$ and $0.025$). 

We run the task on the 4 datasets described in Section \ref{subsec:Datasets} and select the optimal parameters based on the HR@10 performance, given that we observe a strong correlation with NDCG@10 performance. Results (average score over 10 folds) and $95\%$ confidence intervals on the test set are aggregated in Table \ref{tab:freq} ("Fully Optimized SGNS"). To demonstrate the benefit of performing a hyperparameter search, we present the results obtained when SGNS is used with default values as defined in Gensim \cite{gensim} implementation in the "Out-of-the-box SGNS" row. As parameter $\alpha$ is often not tunable in implementations and never discussed in the recommendation setting, we also report the results obtained when optimizing on every hyperparameter but $\alpha$ in the "\textit{Optimized} SGNS" row, in order to isolate the benefit of optimizating over this variable.

To compare the benefit of using recommendation-specific implementations of Prod2vec, we use Meta-Prod2vec \cite{metaprod2vec} on the two music datasets, having artists as side information, and report results in the "Fully optimized Meta-Prod2vec" row for optimized models, and "Meta-Prod2vec \cite{metaprod2vec}" for models trained with the configuration specified in \cite{metaprod2vec}. As Meta-Prod2vec was specifically designed to perform well on a cold-start regime, we also report, in Table \ref{tab:freq2}, results on the cold-start scenario, for pair of (query item, next item) that have zero or less than $3$ co-occurrences in the training set.


\section{Results}

\begin{table*}
  \caption{Next Event Prediction results for music, e-commerce and click-stream datasets with different hyperparameter optimization strategy. The 4 specified hyperparameters  are respectively: window-size, number of epochs, sub-sampling parameter and negative sampling distribution parameter.}
  \label{tab:freq}
  \resizebox{\textwidth}{!}{
  \begin{tabular}{ccccccccc}
    \toprule
    Model&30Music dataset (HR@10)&30Music dataset (NDCG@10)&Deezer dataset (HR@10)&Deezer dataset (NDCG@10)&E-commerce dataset (HR@10)&E-commerce dataset (NDCG@10)& Click-stream dataset (HR@10) & Click-stream dataset (NDCG@10)\\
    \midrule
    Out-of-the-box SGNS &11.16 $\pm$ 0.1&0.099 $\pm$ 0.001&8.13 $\pm$ 0.1&0.061 $\pm$ 0.004&22.21 $\pm$ 0.1&0.159 $\pm$ 0.001 & 3.07 $\pm$ 0.1& 0.018$\pm$ 0.001\\ 
    (hyperparameters: $L,n,t,\alpha$) &(5,5,10$^{-3}$,0.75)&(5,5,10$^{-3}$,0.75)&(5,5,10$^{-3}$,0.75)&(5,5,10$^{-3}$,0.75)&(5,5,10$^{-3}$,0.75)&(5,5,10$^{-3}$,0.75) &(5,5,10$^{-3}$,0.75)&(5,5,10$^{-3}$,0.75)\\[0.5cm]
    \textit{Optimized} SGNS&22.24 $\pm$ 0.1&0.166 $\pm$ 0.001&14.43 $\pm$ 0.1&0.100 $\pm$ 0.001&26.17 $\pm$ 0.1&0.181 $\pm$ 0.001 & 24.14 $\pm$ 0.5 &0.130  $\pm$ 0.003 \\
    (hyperparameters: $L,n,t,\alpha$) &(3,90,10$^{-5}$,0.75)&(3,90,10$^{-5}$,0.75)&(3,90,10$^{-5}$,0.75)&(3,90,10$^{-5}$,0.75)&(3,140,10$^{-3}$,0.75)&(3,140,10$^{-3}$,0.75) &(7,150,10$^{-5}$,0.75)&(7,150,10$^{-5}$,0.75)\\[0.5cm]
    Fully optimized SGNS&\textbf{23.75} $\pm$ 0.1&\textbf{0.174} $\pm$ 0.001&\textbf{15.73} $\pm$ 0.1&\textbf{0.108} $\pm$ 0.001&\textbf{26.34} $\pm$ 0.1&\textbf{0.183} $\pm$ 0.001 & \textbf{26.26} $\pm$ 0.2& \textbf{0.147} $\pm$ 0.002\\
    (hyperparameters: $L,n,t,\alpha$)&(3,110,10$^{-5}$,-0.5)&(3,110,10$^{-5}$,-0.5)&(3,130,10$^{-5}$,-0.5)&(3,130,10$^{-5}$,-0.5)&(3,140,10$^{-3}$,1)&(3,140,10$^{-3}$,1) &(7,150,10$^{-5}$,-1)&(7,150,10$^{-5}$,-1)\\[0.5cm]
    MetaProd2vec \cite{metaprod2vec} & $19.41 \pm 0.2$& $0.142\pm 0.001$&$14.24$ $\pm$ $0.1$&$0.097$ $\pm$ 0.001&-&-&-&-\\
    (hyperparameters: $L,n,t,\alpha$) & ($3,10,10^{-3},0.75$)& ($3,10,10^{-3},0.75$)&($3,10,10^{-3},0.75$)&($3,10,10^{-3},0.75$)\\[0.5cm]
    Fully optimized MetaProd2vec & $20.85 \pm 0.1$& $0.152\pm 0.001$&\textbf{15.62} $\pm$ 0.1&\textbf{0.108} $\pm$ 0.001&-&-&-&-\\
    (hyperparameters: $L,n,t,\alpha$) & ($7,90,10^{-4},-0.5$)& ($7,90,10^{-4},-0.5$)&$(3,150,10^{-4},-0.5)$&$(3,150,10^{-4},-0.5)$\\
  \bottomrule
\end{tabular}
}
\end{table*}

\begin{table}
  \caption{NEP performance (HR@10) in cold-start regime as a function of training frequency of the pair (query item, next item) on 30Music and Deezer datasets. \textcolor{red}{}}
  \label{tab:freq2}
    \resizebox{250pt}{!}{
  \begin{tabular}{ccc}
    \toprule
    Model (dataset)&Pair frequency = 0& Pair frequency < 3\\
    \midrule
    Fully optimized SGNS (30Music)& 8.29 $\pm$ 0.1& \textbf{16.48} $\pm$ 0.1\\
    (hyperparameters: $L,n,t,\alpha$)&(3,110,10$^{-5}$,-0.5)&(3,110,10$^{-5}$,-0.5)\\[0.5cm]
    Fully optimized MetaProd2vec (30Music)& $ \textbf{8.84}  \pm 0.1$& $15.79\pm 0.1$\\
    (hyperparameters: $L,n,t,\alpha$) & ($7,90,10^{-4},-0.5$)& ($7,90,10^{-4},-0.5$)\\[0.2cm]
    \midrule \\
    Fully optimized SGNS (Deezer)&4.52 $\pm$ 0.1&\textbf{9.81} $\pm$ 0.1\\
    (hyperparameters: $L,n,t,\alpha$)&(3,130,10$^{-5}$,-0.5)&(3,130,10$^{-5}$,-0.5)\\[0.5cm]
    Fully optimized MetaProd2vec (Deezer)& $\textbf{5.43}  \pm 0.1$& $\textbf{9.98}\pm 0.1$\\
    (hyperparameters: $L,n,t,\alpha$) & ($3,150,10^{-4},-0.5$)& ($3,150,10^{-4},-0.5$)\\
    \bottomrule
\end{tabular}
}
\end{table}


On the two Music datasets, performing a hyperparameter search roughly doubles the performance (Table \ref{tab:freq}), over using the default values. The best configurations for these two datasets are quasi-identical (same $\alpha$, sub-sampling parameters and window-size), which is possibly a consequence of the observed similarity of count distributions in Section~3. Hyperparameter optimization allows to increase performance by a factor of $10$ for the Click-Stream dataset, and yields substantial performance gains for the E-commerce dataset. 

\begin{figure}[htb]
    \includegraphics[width=1\linewidth]{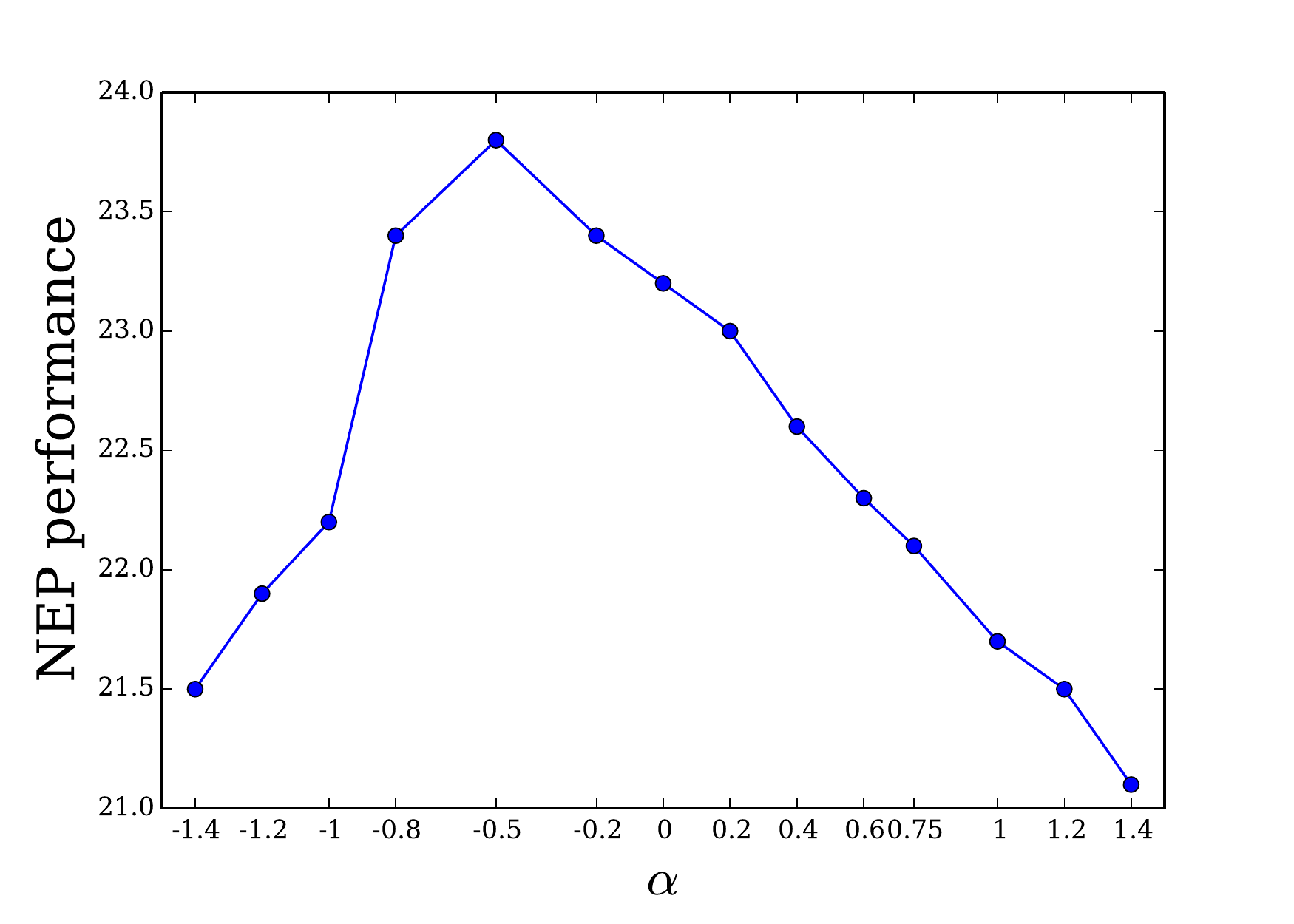}
    \caption{Best final performance of SGNS as a function of negative sampling parameter $\alpha$ for 30Music dataset.}
\label{fig4}
\end{figure}

Interestingly, for all datasets, the marginal benefit of including $\alpha$ in the hyperparameter search is significant in terms of final performance. This is illustrated in Figure~\ref{fig4}, where we select the best performing configurations for different $\alpha$ values and plot the NEP performance for 30Music dataset. The original $\alpha=0.75$, optimal on linguistic tasks, is clearly suboptimal on this recommendation setting. The optimal hyperparameter $\alpha$ is $-0.5$, such that the optimal negative sampling distribution is one more likely to sample unpopular items as negative examples. For the Deezer and Click-Stream datasets, the optimal $\alpha$ is also negative.



We observe that Meta-Prod2vec \cite{metaprod2vec} can also benefit from a hyperparameter optimization, with once again a negative $\alpha$. However, we also note that it is outperformed by an optimized SGNS on 30Music and on par on Deezer dataset (Table \ref{tab:freq}). On the cold-start regime, results indicates that, once optimized, MetaProd2vec is on par with SGNS (Table \ref{tab:freq2}). Hence, it might be worth optimizing standard methods before moving to more specialized methods.

Results confirm that the optimal choice hyperparameters for SGNS are data-dependent and task-dependent, and that, for the given datasets and the considered task (NEP), it is highly valuable in terms of final performance to simultaneously optimize the hyperparameters.
Especially, the optimal negative sampling distribution clearly differs from the one proven to be optimal for linguistic tasks \cite{word2vec2,levy2015improving}, and optimizing over this additional variable yields significant improvements. As a negative $\alpha$ leads to sample more often unpopular items for negative samples, and positive samples are necessarily more popular by construction, the algorithm is made to better distinguish between items of different order of popularity. This correlates with the observation that items tend to share the same order of popularity within a music streaming session.


\section{Conclusion}

Developed first for NLP, SNGS generate words embeddings that help achieving state of the art performance in semantic similarity and analogy tasks. Previous work shows that it can be directly applied to sequences of items to generate item embeddings useful for recommendations. Interestingly, while NLP data and tasks differ in their structure and goals from those of recommendation, the hypothesis behind some of the parameters of the algorithms are barely discussed, nor their default fixed values revised. We show that using different values for some hyperparameters, namely negative sampling distribution, number of epochs, subsampling parameter and window-size, leads to significantly better performances on classical evaluation tasks on 4 recommendation datasets. While performing a hyperparameter search for each different types of data and tasks in a real-life recommendation setting can be time consuming and computationally costly, we stress out the benefits of having better item embeddings to better distinguish, cluster and classify content, which can lead to substantial gains in demanding industries, such as on-demand music streaming services, where a few bad recommendations can quickly lead a user to leave the service.

Comparing several recommendation datasets on the same task, we observe that different data distributions result in different optimal hyperparameter values. The homogeneity of popularity between items of a same sequence, the shape of the popularity distribution, or the heterogeneity of the items in the catalog have a direct impact on the task evaluation. We have yet to find if and how we can induce the optimal hyperparameter values from the structure of the data. This could have a strong impact on improving the current results with SNGS applied to recommendation.


%

\bibliographystyle{ACM-Reference-Format}
\bibliography{sample-bibliography} 

\end{document}